\documentclass[preprint]{revtex4}
\DeclareRobustCommand{\baselinestretch{2}}

\usepackage{graphicx}

\begin{document}

\title{Electromagnetically induced absorption and transparency
in magneto-optical resonances in elliptically polarized field}

\author{D.V. Brazhnikov}
\affiliation{Institute of Laser Physics, Siberian Branch of Russian Academy of Sciences,
Novosibirsk 630090, Russia}
\author{A.V. Taichenachev}
\affiliation{Novosibirsk State University, Novosibirsk 630090, Russia}
\author{A.M. Tumaikin}
\affiliation{Institute of Laser Physics, Siberian Branch of Russian Academy of Sciences,
Novosibirsk 630090, Russia}
\author{V.I. Yudin}
\affiliation{Institute of Laser Physics, Siberian Branch of Russian Academy of Sciences,
Novosibirsk 630090, Russia}

\begin{abstract}
Using a $1 \to 2$ transition as an analytically tractable model, we discuss in detail
magneto-optical resonances of both EIA (electromagnetically induced absorption) and EIT
(electromagnetically induced transparency) types in the Hanle configuration. The analysis is made
for arbitrary rate of depolarizing collisions in the excited state and arbitrary elliptical field
polarization. The obtained results clearly show that the main reason for the EIA sub-natural
resonance is the spontaneous transfer of anisotropy from the excited level to the ground one. In
the EIA case we predict the negative structures in the absortpion resonance at large field
detuning. The role of the finite atom-light interaction time is briefly discussed. In addition we
study non-trivial peculiarities of the resonance lineshape related to the velocity spread in a gas.
\end{abstract}


\maketitle 

\section{Introduction}
Nonlinear interference effects, based on atomic coherence \cite{scullyQO}, have attracted in the
past decade a growing attention. They have found numerous applications in nonlinear optics
\cite{NLO} - \cite{Wang}, nonlinear high-resolution spectroscopy \cite{DSres} - \cite{Levine},
high-precision metrology (atomic clocks and magnetometers) \cite{DSmetr} - \cite{Jansses}, laser
cooling \cite{VSCPT,Aspect}, atom optics and interferometry \cite{AT} - \cite{Featonby}, and
quantum information processing \cite{DSPol} - \cite{van der Wal}.

To date the most investigated phenomenon is electromagnetically induced transparency (EIT)
\cite{EIT}. As a matter of fact, EIT means a broad array of nonlinear effects in which the
absorption is significantly reduced due to nonlinear interference of electromagnetic waves. In
particular, in certain atomic systems EIT is related to dark states and coherent population
trapping (CPT) \cite{CPT,Arimondo}.

The opposite phenomenon of electromagnetically induced absorption (EIA), first observed by Akulshin
et al. in 1998 \cite{akulshin98}, is much less studied. After 1998 several groups have performed
experiments on observations of EIA and related effects in different atomic systems and under
different conditions \cite{EIA} - \cite{Andreeva}. Both two-photon resonances in a bichromatic
light field and magneto-optical resonances in the Hanle configuration have been explored.

The first theoretical explanation of physical origins of EIA was given in
\cite{taich99,Taichenachev}, where, using a simple analytically tractable model of a four-level
$N$-system as an example, we have shown that EIA is due to the transfer of low-frequency atomic
coherence between interacting levels in the course of spontaneous radiative transitions.
Subsequently, this concept has been confirmed in experiments and by numerical calculations
\cite{failache03}, and it has been further developed in \cite{WGtop}. Summarizing and
generalizing, we can say that EIA  can be observed on resonant atomic transitions of $F_g=F \to
F_e=F+1$ type with degenerate ground state ($F>0$), and its physical origin is the spontaneous
transfer of the light-induced anisotropy (atomic coherence or/and population difference). As the
result, under two-photon resonance conditions atoms mainly populate Zeeman sub-states most coupled
to the light field. If the spontaneous transfer of anisotropy is absent, i.e., the ground state is
repopulated isotropically, atomic populations are distributed inversely proportional to the
optical depopulation rates, which leads to the EIT-type resonances.

It is worth to note that the same physical reason (the spontaneous transfer of anisotropy) is
responsible for the "anomalous" sign of nonlinear magneto-optical rotation of linearly polarized
light on $F \to F+1$ transitions. This effect has been observed earlier in 1990 \cite{weis90}, and
its theoretical explanation has been given by Kanorsky et al. \cite{kanorsky93} on the base of the
perturbation theory method developed in \cite{yudin88}. The difference in  signs of the
ground-state quadrupole moments, which govern the sign of nonlinear magneto-optical rotation, for
transitions $F \to F+1$ from one hand, and for $F \to F$ and $F \to F-1$ transitions from the other
hand has been discussed in \cite{kazantsev84}.

The present paper is devoted to a detailed theoretical study of EIA/EIT magneto-optical resonances
in the Hanle configuration. We explore the realistic but analytically solvable model of a $F_g=1
\to F_e=2$ atomic transition. Note that this transition is the simplest transition (from the $F \to
F+1$ class), where the light-induced quadrupole moment, precessing in a magnetic field orthogonal
to the polarization ellipse plane, can appear in the ground state. Apart from the radiative
relaxation, the excited-state depolarization due to collisions with a buffer gas is taken into
account, which allows us to consider a gradual transition from EIA to EIT with the increase of the
depolarization rate. It should be noted, in the majority of works on the magneto-optical variant
of EIA/EIT resonance, except for the papers \cite{milner99} - \cite{budker03}, the linearly
polarized radiation was used or considered. Here we investigate a general case of the resonance
interaction of atoms with elliptically polarized light. We find several new features of the
magneto-optical spectra connected with the light ellipticity. The treatment is carried out for
atoms at rest (homogeneous broadening) as well as for atomic gas under conditions of the Doppler
broadening. The role of the finite atom-light interaction time is briefly discussed.

\section{Problem statement}
\begin{figure}[h]\centerline{\scalebox{0.5}{\includegraphics{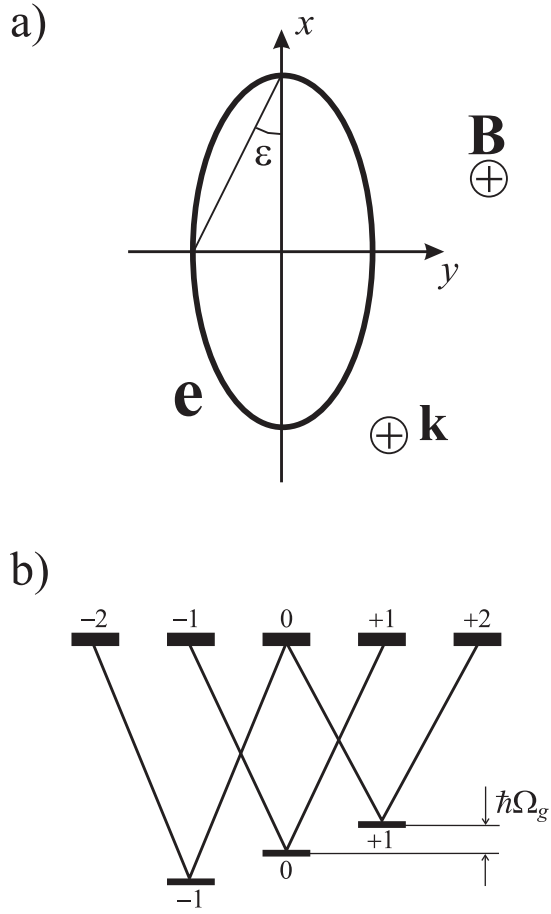}}}
\caption{(a) Mutual orientation of the polarization ellipse and the magnetic field, and (b) the
scheme of light-induced transitions.} \label{scheme}
\end{figure}

We consider the resonance interaction of atoms, having the total angular momenta $F_g$ in the
ground state and $F_e$ in the excited state, with a monochromatic light field ${\bf E}({\bf r},t)$
in the presence of a static magnetic field ${\bf B}$. More specifically, the field configuration is
an elliptically polarized running wave:
\begin{equation}\label{Et}
{\bf E}({\bf r},t)=E\,{\bf e}\,\exp\{-i(\omega t-{\bf kr})\}+c.c.,
\end{equation}
where $E$ is the complex amplitude, and ${\bf e}$ is the unit polarization vector. In the
coordinate frame associated with the wave (${\bf e}_z$ is directed along ${\bf k}$, ${\bf e}_{x,y}$
-- along the polarization ellipse semiaxes) the polarization vector can be written as
\begin{equation} \label{e}
{\bf e}={\bf e}_x\cos(\varepsilon)+i{\bf e}_y\sin(\varepsilon)={\bf
e}_{+1}\cos(\varepsilon-\pi/4)+{\bf e}_{-1}\sin(\varepsilon-\pi/4)\;,
\end{equation}
with ${\bf e}_{\pm1}=\mp({\bf e}_x\pm i{\bf e}_y)/\sqrt{2}$ the corresponding spherical orths, and
$\tan(\varepsilon)$ equal to the ratio of the ellipse semiaxes (see in Fig.~\ref{scheme}.a).

The atom-light interaction Hamiltonian in the dipole and in the rotating-wave approximations has
the form:
\begin{equation}
\widehat{H}_{D-E}=-\widehat{\bf d}{\bf E}=\hbar \kappa \widehat{V} + h.c. \;,
\end{equation}
where $\kappa = -\langle F_e||\widehat{d}||F_g\rangle E/\hbar$ is the coupling constant (Rabi
frequency), and, according to the Wigner-Eckart theorem,
\begin{equation} \label{V^E}
\widehat{V}=\widehat{\bf T}^{eg} \,{\bf e}\,,
\end{equation}
with the Wigner vector operator defined through Clebsch-Gordan coefficients:
\begin{equation} \label{T1}
\widehat{T}^{ab}_{q}=\sum_{m_a,m_b} C^{F_a\,m_a}_{F_b\,m_b\,;1\,q}\, |F_a,m_a\rangle\,\langle
F_b,m_b|\,,
\end{equation}
the indices $a$ and $b$ can take values $e,\,g$. The interaction with a weak static field ${\bf B}$
is described by the Hamiltonian
\begin{equation} \widehat{H}_B=-\widehat{\mathbf{\mu}}{\bf B}
=\sum_{a=e,g}\hbar\,\Omega_a\,\widehat{F}^{(a)}_b\;,
\end{equation}
where the operator
\begin{equation} \label{V^H}
\widehat{F}^{(a)}_b=\sqrt{F_a(F_a+1)}\, \widehat{\bf T}^{aa}\,{\bf b}\;
\end{equation}
are the projection of the total angular momentum operator of the level $(a)$ on
the direction ${\bf b}$; $\hbar\Omega_a=\mu_B g_a B$ is the Zeeman splitting
between adjacent substates of the level $(a)$ with $\mu_B$ the Bohr magneton
and $g_a$ the Lande $g$-factor; ${\bf b}={\bf B}/B$ is a unit vector directed
along the magnetic field ${\bf B}$.

All operators are represented as matrices on the Zeeman basis of the ground and
excited levels, with states $\{|F_{g},m _{g}\rangle \}$ and $\{|F_{e},m
_{e}\rangle \}$. The atomic density matrix $\widehat{\rho }$ can be separated
in four matrix blocks, where the matrices $\widehat{\rho }_{gg}$ and $\widehat{
\rho }_{ee}$ are the density submatrices for the ground and excited state, and
the off-diagonal blocks $\widehat{\rho }_{eg}$ and $\widehat{\rho }_{ge}$
describe the optical coherences. In the rotating-wave approximation the fast
time-space dependence of the kinetic equation can be removed by introducing the
transformed optical coherences as
\begin{equation}
\widehat{\rho }_{eg}=\widehat{\overline{\rho }}_{eg}\,\exp\{-i(\omega t-{\bf kr})\}\;\;\;\;%
\widehat{\rho }_{ge}=\widehat{\overline{\rho }}_{ge}\,\exp\{i(\omega t-{\bf kr})\}. \label{average}
\end{equation}
Then, the generalized optical Bloch equations (GOBE) take the form
\begin{equation} \label{eg}
(\gamma_{eg}-i\delta_v)\widehat{\overline{\rho }}_{eg}=
-i\,\kappa\,\left[\widehat{V}
\widehat{\rho}_{gg}-\widehat{\rho}_{ee}\widehat{V}\right]
-i\,\left[\Omega_e\,\widehat{F}^{(e)}_b \widehat{\overline{\rho
}}^{eg}-\widehat{\overline{\rho }}_{eg}\,\Omega_g\,\widehat{F}^{(g)}_b\right];
\end{equation}
\begin{equation} \label{ge}
(\gamma_{eg}+i\delta_v) \widehat{\overline{\rho }}_{ge}=
-i\,\kappa\,\left[\widehat{V}^{\dagger}
\widehat{\rho}_{ee}-\widehat{\rho}_{gg}\widehat{V}^{\dagger}\right]
-i\,\left[\Omega_g\,\widehat{F}^{(g)}_b \widehat{\overline{\rho
}}^{ge}-\widehat{\overline{\rho }}_{ge}\,\Omega_e\,\widehat{F}^{(e)}_b\right];
\end{equation}
\begin{equation} \label{ee}
(\Gamma+\gamma_r)\widehat{\rho}_{ee}
+\widehat{\gamma}_{coll}\{\widehat{\rho}_{ee}\} = -i\,\kappa\,\left[\widehat{V}
\widehat{\overline{\rho }}_{ge}-\widehat{\overline{\rho
}}^{eg}\widehat{V}^{\dagger}\right] -i\,\Omega_e\,\left[\widehat{F}^{(e)}_b,\,
\widehat{\rho}_{ee}\right];
\end{equation}
\begin{equation} \label{gg}
\Gamma\left(\widehat{\rho}_{gg}- \widehat{\rho}_{gg}^{(0)}\right)=\widehat{\cal
G}\{\widehat{\rho}_{ee} \}-i\,\kappa\,\left[\widehat{V}^{\dagger}
\widehat{\overline{\rho }}_{eg}-\widehat{\overline{\rho
}}^{ge}\widehat{V}\right] -i\,\Omega_g\,\left[\widehat{F}^{(g)}_b,\,
\widehat{\rho}_{gg}\right] \;,
\end{equation}
where $[\,,\,]$ indicates a commutator, $\delta_v=\omega-\omega_{eg}-kv$ is the detuning with
account for the Doppler shift $kv$ for a moving atom, $\gamma_{eg}$ is the dephasing rate, in
Eq.(\ref{ee}) the parameter $\gamma_r$ is the radiative relaxation rate, the operator
$\widehat{\gamma}_{coll}\{\widehat{\rho}_{ee}\}$ describes the collisional depolarization in the
excited state, the rate $\Gamma$ describes relaxation of atoms to the isotropical distribution
$\widehat{\rho}_{gg}^{(0)}=\widehat{\Pi}_g/(2F_g+1)$ (the operator
$\widehat{\Pi}_a=\sum_{m_a}|F_a,m_a\rangle \langle F_a,m_a|$ projects on the given energy level
$(a)$) outside the light beam due to either atomic free flight through the beam or diffusion to
the walls. The operator
\begin{equation} \label{spont}
\widehat{\cal G}\{\widehat{\rho}_{ee}\}=\beta\,\gamma_r\, \sum_{q=0,\pm1} \widehat{T}^
{eg\,\dagger}_{q}\widehat{\rho}_{ee}\: \widehat{T}^{eg}_{q}
\end{equation}
in the right-hand side of Eq.(\ref{gg}) corresponds to the repopulation of the ground state due to
the spontaneous radiative transitions. This process includes the transfer of the Zeeman-substate
populations (terms with diagonal matrix elements $\rho_{m_em_e}$) as well as the transfer of the
Zeeman coherence (terms with off-diagonal matrix elements $\rho^{ee}_{m_em'_e}$ at $m_e\ne m'_e$).
In general, it should be viewed as the spontaneous transfer of the total population and of the
Zeeman anisotropy between working levels. The coefficient $\beta\le 1$ governs the branching ratio
in the course of the spontaneous decay from the excited level $F_e$ to the lower level $F_g$. When
$\beta=1$ the transition $F_g\rightarrow F_e$ is closed, i.e. the total population is conserved
($\mbox{Tr}\{\widehat{\rho}_{gg}\}+\mbox{Tr}\{\widehat{\rho}_{ee}\} =1$). Hereafter the symbol
$\mbox{Tr}\{...\}$ means the trace operation over internal degrees of freedom.

We assume the property
\begin{equation} \label{conserv}
\mbox{Tr}\left\{\widehat{\gamma}_{coll}\{\widehat{\rho}_{ee}\}\right\}= 0\;.
\end{equation}
If collisions are negligible, we have
\begin{equation} \label{sp}
\widehat{\gamma}_{coll}=0\,,\quad \gamma_{eg}=\gamma_r/2+\Gamma\,.
\end{equation}
The collisional relaxation term
$\widehat{\gamma}_{coll}\{\widehat{\rho}_{ee}\}$ has the simplest form
\begin{equation} \label{coll}
\widehat{\gamma}_{coll}\{\widehat{\rho}_{ee}\}=\gamma_1
\left[\widehat{\rho}_{ee}- \widehat{\Pi}_e
\mbox{Tr}\left\{\widehat{\rho}_{ee}\right\}/(2F_e+1) \right]
\end{equation}
in the model case, when all the multipole moments relax due to collisions with the same rate
$\gamma_1$ (apart from the total population, which is conserved according to Eq.(\ref{conserv})).

More specifically, we will study just one closed ($\beta=1$) transition
$F_g=1\to F_e=2$. This transition is realistic and simultaneously sufficiently
simple, allowing analytical description of the EIA/EIT effects. Let the static
magnetic field is directed orthogonal to the polarization ellipse, i.e. along
the $z$ axis (see in Fig.~\ref{scheme}.a). The corresponding scheme of the
light-induced transitions is shown in Fig.~\ref{scheme}.b. As a spectroscopic
signal we consider the total excited-state population as a function of the
magnetic field amplitude $B$ (Hanle-type spectroscopy)
\begin{equation}\label{ne}
\pi_e=\mbox{Tr}\left\{\widehat{\rho}_{ee}\right\}\,.
\end{equation}
This signal is proportional to the total fluorescence and to the total light
absorption in optically thin media. We investigate the influence of the
radiative relaxation operator (\ref{spont}) in combination with the collisional
depolarization operator $\widehat{\gamma}_{coll}\{\widehat{\rho}_{ee}\}$ on the
lineshape of the resonance described by Eq.(\ref{ne}).

We will be interested in sub-natural width structures ($\Omega_{e,g} \ll \gamma_r$), which appear
in nonlinear spectra in the low-saturation limit, when the light field is sufficiently weak:
\begin{equation} \label{lowsat}
S=\frac{|\kappa|^2}{\gamma_{eg}^2+\delta_v^2} \ll 1 \;.
\end{equation}
With these approximations, eliminating the optical coherences, we arrive at the
following closed set of equations for the excited-state and ground-state
density matrices:
\begin{equation} \label{goberedE}
(\Gamma+\gamma_r)\widehat{\rho}_{ee} +\gamma_1 \left[\widehat{\rho}_{ee}-
\widehat{\Pi}_e \,\mbox{Tr}\left\{\widehat{\rho}_{ee}\right\}/(2F_e+1) \right]
= 2\gamma_{eg}\,S\, \widehat{V}\widehat{\rho}_{gg}\widehat{V}^{\dagger} \;,
\end{equation}
\begin{equation} \label{goberedG}
\Gamma\left(\widehat{\rho}_{gg}-
\widehat{\rho}_{gg}^{(0)}\right)+i\Omega_g[\widehat{F}^{(g)}_b,\widehat{\rho}_{gg}]
=-\left\{(\gamma_{eg}+i\delta_v)\,S\,
\widehat{V}^{\dagger}\widehat{V}\widehat{\rho}_{gg}+h.c.\right\} + \gamma_r\,
\sum_{q=0,\pm1} \widehat{T}^ {eg\,\dagger}_{q}\,\widehat{\rho}_{ee}\,
\widehat{T}^{eg}_{q}\;,
\end{equation}
where the collision relaxation operator is taken in the form (\ref{coll}).

\section{Solution for atom at rest}
Consider first the total excited-state population $\pi_e$ of an atom at rest ($\mathbf{v}=0$), when
the detuning $\delta=\omega-\omega_{eg}$. The solution for a moving atom with the given velocity
$\mathbf{v}$ can be easily derived from the expressions below by the substitution $\delta \to
\delta_v$. As is seen from Eq.(\ref{goberedG}) and Fig.1.b, the coherence between just two Zeeman
substates ($|F_g=1,\,m_g=\pm1\rangle$) is sensitive to the magnetic field. Thus, as it has been
shown in \cite{knappe02}, $\pi_e$ (as well as any spectroscopic signal) is a quotient of
polynomials of second order in $\Omega_g$:
\begin{equation} \label{quotient}
\frac{\pi_e}{\pi_e^{(0)}} =\frac{\sum_{i=0}^{2}
\mathcal{N}_i(\Delta,\widetilde{\gamma}_1,\widetilde{\Gamma},\varepsilon)\,\Omega^i}{\sum_{k=0}^{2}
\mathcal{D}_k(\Delta,\widetilde{\gamma}_1,\widetilde{\Gamma},\varepsilon)\,\Omega^k}\;,
\end{equation}
where $\pi_e^{(0)} = 2\,\gamma_{eg}\,S/(\gamma_r+\Gamma)$ corresponds to the linear absorption of
unpolarized atoms, $\Omega=\Omega_g/(\gamma_{eg}\,S)$, $\Delta = \delta/\gamma_{eg}$,
$\widetilde{\gamma}_1 = \gamma_1/\gamma_r$, $\widetilde{\Gamma}=\Gamma/(\gamma_{eg}\,S)$, and
$\varepsilon$ is the ellipticity parameter of the light wave. The numerator and denominator can be
expanded in $\Delta$ powers:
\begin{equation} \label{numerator}
\mathcal{N}_i
=\sum_{j=0}^{2}\mathcal{N}_{ij}(\widetilde{\gamma}_1,\widetilde{\Gamma},\varepsilon)\,\Delta^j \;,
\end{equation}
\begin{equation} \label{denomenator}
\mathcal{D}_k
=\sum_{l=0}^{2}\mathcal{D}_{kl}(\widetilde{\gamma}_1,\widetilde{\Gamma},\varepsilon)\,\Delta^l \;.
\end{equation}
The coefficients $\mathcal{N}_{ij}$ and $\mathcal{D}_{kl}$ with $(i+j)$ and $(k+l)$ odd numbers are
equal to zero due to symmetry reasons. For the sake of brevity, here we give explicit analytical
expressions for  nonzero coefficients $\mathcal{N}_{ij}$ and $\mathcal{D}_{kl}$ in the limiting
case, when the power broadening $\gamma_{eg}\, S$ dominates over the transit-time broadening
$\Gamma$, i.e. at $\widetilde{\Gamma}=0$:
\begin{eqnarray} \label{Nij}
\mathcal{N}_{00} &=& \,\left( 5 + 7\,\widetilde{\gamma}_1 \right)
\,\left[ 25 + 115\,\widetilde{\gamma}_1 +
172\,\widetilde{\gamma}_1^2 + 84\,\widetilde{\gamma}_1^3 -
    c^2\,\left( 15 + 9\,\widetilde{\gamma}_1 - 76\,\widetilde{\gamma}_1^2 -
    84\,\widetilde{\gamma}_1^3 \right) - 4\,c^4\,\widetilde{\gamma}_1 \right]\nonumber\\
\mathcal{N}_{02} &=& -\,4{\left( 1 + \widetilde{\gamma}_1 \right)
}^2\,\left[- 25\,\left( 5 + 16\,\widetilde{\gamma}_1 +
12\,\widetilde{\gamma}_1^2 \right) + c^2\,\left( 175 +
320\,\widetilde{\gamma}_1 - 12\,\widetilde{\gamma}_1^2 \right) -
12\,c^4\,\left( 5 - 4\,\widetilde{\gamma}_1 -
24\,\widetilde{\gamma}_1^2 \right)\right]\nonumber\\
\mathcal{N}_{11} &=&160\,s\,{\left( 1 + \widetilde{\gamma}_1
\right) }^2\,\left[ 15 + 48\,\widetilde{\gamma}_1 +
36\,\widetilde{\gamma}_1^2 -
    c^2\,\left(8 - 6\,\widetilde{\gamma}_1 - 36\,\widetilde{\gamma}_1^2 \right)  \right]  \nonumber\\
\mathcal{N}_{20} &=& 48\,{\left( 1 + \widetilde{\gamma}_1 \right)
}^2\,\left[ 12\,\left( 1 + 2\,\widetilde{\gamma}_1 \right) \,
     \left( 5 + 6\,\widetilde{\gamma}_1 \right)  -
     5\,c^2\,\left( 7 - 5\,\widetilde{\gamma}_1 - 30\,\widetilde{\gamma}_1^2 \right)\right] \;,
\end{eqnarray}
\begin{eqnarray} \label{Dkl}
\mathcal{D}_{00} &=& \,\left( 5 + 7\,\widetilde{\gamma}_1 \right)
\,\left[ \left( 5 + 7\,\widetilde{\gamma}_1 \right) \,
     \left( 5 + 32\,\widetilde{\gamma}_1 + 36\,\widetilde{\gamma}_1^2 \right)  -
    4\,c^2\,\left( 2 + 5\,\widetilde{\gamma}_1 - 8\,\widetilde{\gamma}_1^2 -
    14\,\widetilde{\gamma}_1^3 \right)  \right] \nonumber\\
\mathcal{D}_{02} &=& -\,4{\left( 1 + \widetilde{\gamma}_1 \right)
}^2\,\left[ -
    25\,\left( 5 + 32\,\widetilde{\gamma}_1 + 36\,\widetilde{\gamma}_1^2 \right)  +
    4\,c^2\,\left( 35 + 194\,\widetilde{\gamma}_1 + 166\,\widetilde{\gamma}_1^2 \right)-32\,c^4\,\left( 1
    + \widetilde{\gamma}_1 - 6\,\widetilde{\gamma}_1^2 \right)\right]\nonumber\\
\mathcal{D}_{11} &=& 160\,s\,{\left( 1 + \widetilde{\gamma}_1
\right) }^2\,\left[
    3\,\left( 5 + 32\,\widetilde{\gamma}_1 + 36\,\widetilde{\gamma}_1^2 \right)-c^2\,\left( 4 +
2\,\widetilde{\gamma}_1 - 24\,\widetilde{\gamma}_1^2 \right)\right] \nonumber\\
\mathcal{D}_{20} &=& 192\,{\left( 1 + \widetilde{\gamma}_1 \right)
}^2\,\left[
    3\,\left( 5 + 32\,\widetilde{\gamma}_1 + 36\,\widetilde{\gamma}_1^2 \right) - c^2\,\left(4 -
25\,\widetilde{\gamma}_1^2 \right)\right]\;,
\end{eqnarray}
where $c = \cos(2\varepsilon)$ and  $s = \sin(2\varepsilon)$.

It is convenient for analysis to present the lineshape (\ref{quotient})  in the
form of generalized Lorentzian \cite{knappe02,taich03}:
\begin{equation} \label{genLor}
\pi_e =
A\,\frac{w^2}{(\Omega_g-\Omega_0)^2+w^2}+B\,\frac{w\,(\Omega_g-\Omega_0)}{(\Omega_g-\Omega_0)^2+w^2}+C
\;,
\end{equation}
where all the parameters are expressed through the coefficients $\mathcal{N}_i$ and
$\mathcal{D}_k$ (see equations (\ref{numerator}-\ref{Dkl})) in the following way. The amplitude of
the symmetric part
\[
\frac{A}{\pi_e^{(0)}}=\frac{2\,(2\mathcal{N}_0\,\mathcal{D}_2^2+\mathcal{N}_2\,\mathcal{D}_1^2-
\mathcal{N}_1\,\mathcal{D}_1\,\mathcal{D}_2-2\,\mathcal{N}_2\,\mathcal{D}_0\,\mathcal{D}_2)}{\mathcal{D}_2\,(4\,\mathcal{D}_0
\,\mathcal{D}_2-\mathcal{D}_1^2)} \;,
\]
the amplitude of antisymmetric part
\[
\frac{B}{\pi_e^{(0)}}=\frac{2\,(\mathcal{N}_1\,\mathcal{D}_2-\mathcal{N}_2\,\mathcal{D}_1)}{\mathcal{D}_2\,\sqrt{4\,\mathcal{D}_0
\,\mathcal{D}_2-\mathcal{D}_1^2}} \;,
\]
the background
\[
\frac{C}{\pi_e^{(0)}} = \frac{\mathcal{N}_2}{\mathcal{D}_2} \;,
\]
the resonance position
\[
\frac{\Omega_0}{\gamma_{eg}\,S} =
-\frac{\mathcal{D}_1}{2\,\mathcal{D}_2} \;,
\]
and the resonance width
\[
\frac{w}{\gamma_{eg}\,S} = \frac{\sqrt{4\,\mathcal{D}_0
\,\mathcal{D}_2-\mathcal{D}_1^2}}{2\,|\mathcal{D}_2|} \;.
\]

It is important that $B$ and $\Omega_0$ are proportional to the detuning $\Delta$ and to the degree
of circular polarization $s=\sin(2\,\varepsilon)$. In other words, the spectroscopic signal is
symmetrical with respect to zero of the magnetic field either at zero detuning or in the case of
linear polarization. In these symmetric cases the sign of $A$ depends on the depolarization rate
$\widetilde{\gamma}_1$. For instance, when $\varepsilon=0$
\[
\frac{A}{\pi_e^{(0)}}=\frac{3\,\left( 5 - 2\,\widetilde{\gamma}_1
\right) \,\left( 1 + 2\,\widetilde{\gamma}_1 \right) }
  {4\,\left( 187 + 565\,\widetilde{\gamma}_1  + 418\,{\widetilde{\gamma}_1 }^2
  \right)}\,
\]
and the sign of the resonance is changed, EIA is transformed into EIT, at $\widetilde{\gamma}_1 =
5/2$ independently of the detuning. In the other case $\Delta=0$ the sign-reversal point weakly
depends on the ellipticity $\varepsilon$ and it lies between $\widetilde{\gamma}_1 = 2$ for
$\varepsilon \to \pm\,\pi/4$ and $\widetilde{\gamma}_1 = 5/2$ for linearly polarized light.

\begin{figure}[h]\centerline{\scalebox{0.8}{\includegraphics{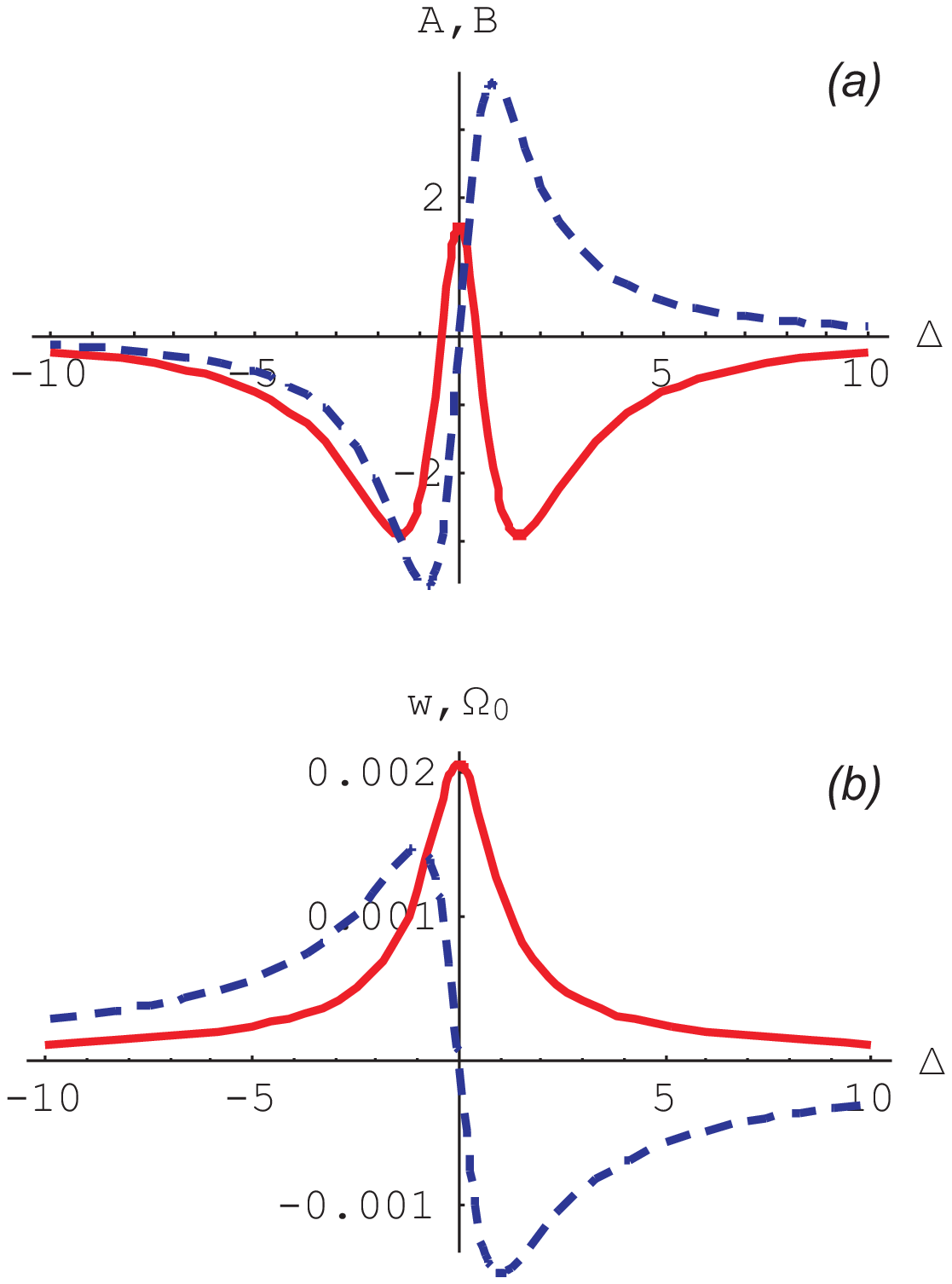}}}
\caption{The lineshape parameters versus $\Delta$. (a) $A$ (solid line) and $B$ (dashed line) in
arbitrary units, and (b) $w$ (solid line) and $\Omega_0$ (dashed line) in $\gamma_{eg}$ units. The
case of the pure radiative relaxation $\gamma_1 = 0$. Other parameters $\varepsilon = \pi/8$,
$\Gamma =0$, $\kappa = 0.1\,\gamma_{eg}$.} \label{LLparG0gk0pio8}
\end{figure}

\begin{figure}[h]\centerline{\scalebox{0.8}{\includegraphics{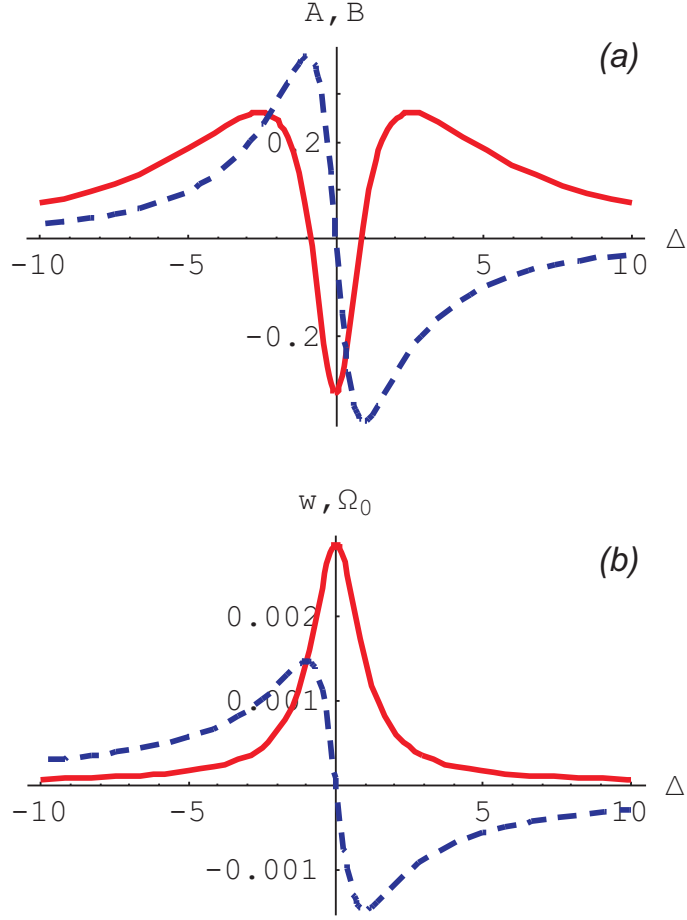}}}
\caption{The lineshape parameters versus $\Delta$. (a) $A$ (solid line) and $B$
(dashed line) in arbitrary units, and (b) $w$ (solid line) and $\Omega_0$
(dashed line) in $\gamma_{eg}$ units. The case of the total excited-state
depolarization $\gamma_1 = 10$. Other parameters $\varepsilon = \pi/8$, $\Gamma
=0$, $\kappa = 0.1\,\gamma_{eg}$.} \label{LLparG0gk10pio8}
\end{figure}

\begin{figure}[h]\centerline{\scalebox{0.8}{\includegraphics{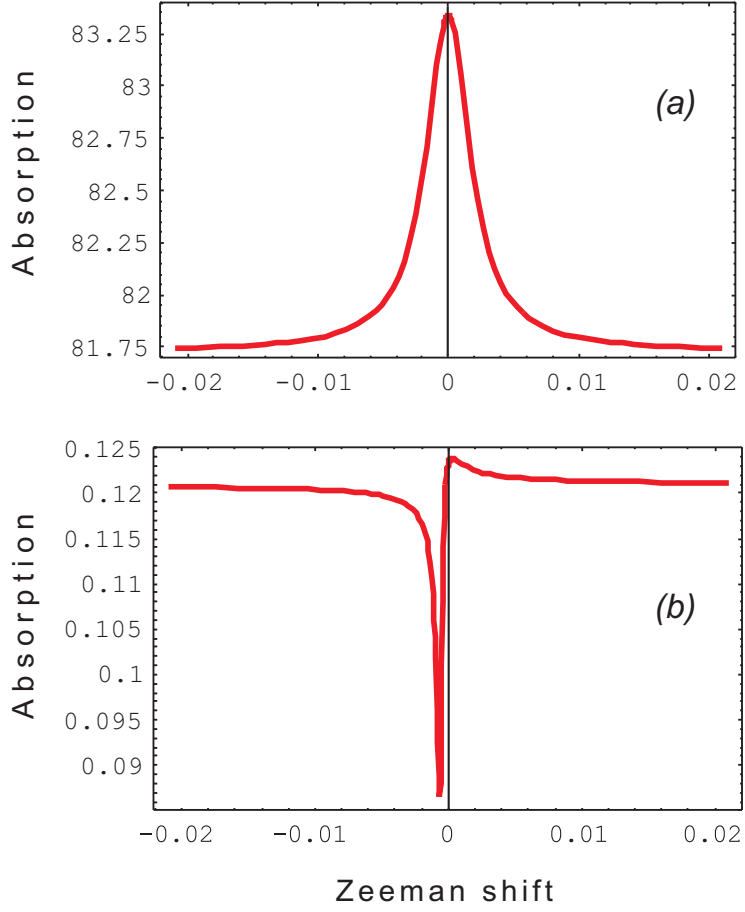}}}
\caption{The absorption resonance lineshapes in the EIA case $\gamma_1 = 0$.
(a) The total excited-state population in arbitrary units versus the
ground-state Zeeman shift in  $\gamma_{eg}$ units at $\Delta = 0$; (b) the same
but $\Delta =5\,\gamma_{eg}$. Other parameters $\varepsilon = \pi/8$, $\Gamma
=0$, $\kappa = 0.1\,\gamma_{eg}$.} \label{lineshapesEIA}
\end{figure}

In the general case, when $\varepsilon \neq 0$ and $\Delta \neq 0$, the signal is asymmetric and
its position is shifted with respect to the zero magnetic field point ($\Omega_g = 0$). At
$\varepsilon=\pi/8$ the dependence of the lineshape parameters $A$, $B$, $w$, and $\Omega_0$ on
the detuning $\Delta$ is shown in Fig.~\ref{LLparG0gk0pio8} and Fig.~\ref{LLparG0gk10pio8} in two
opposite cases. Fig.~\ref{LLparG0gk0pio8} corresponds to the pure radiative relaxation $\gamma_1 =
0$, when the transfer of anisotropy is maximal, while at $\gamma_1 = 10\, \gamma_r$ used in
Fig.~\ref{LLparG0gk10pio8} the collisional depolarization of the excited state is almost complete.
As is seen from these figures, the amplitudes of symmetric $A$ and antisymmetric $B$ parts are
comparable. It should be also noted that the amplitude $A$ changes sign when the detuning $\Delta$
increases. Such a behaviour in the EIT case was well-known in the simplest model of a three-level
$\Lambda$ system \cite{rautian79}. This effect is usually related to the Raman absorption peak in
the probe field spectra \cite{rautian79} - \cite{Lounis}. In the EIA case at large detunings we
see a negative structure in the absorption resonance (Fig.~\ref{lineshapesEIA}.b), which can be
also attributed to the Raman scattering. This structure is asymmetric and significantly narrower
than the EIA resonance at $\Delta = 0$ (Fig.~\ref{lineshapesEIA}.a). Such a Raman transparency
resonance has not been discussed previously to the best of our knowledge.

\begin{figure}[h]\centerline{\scalebox{1.0}{\includegraphics{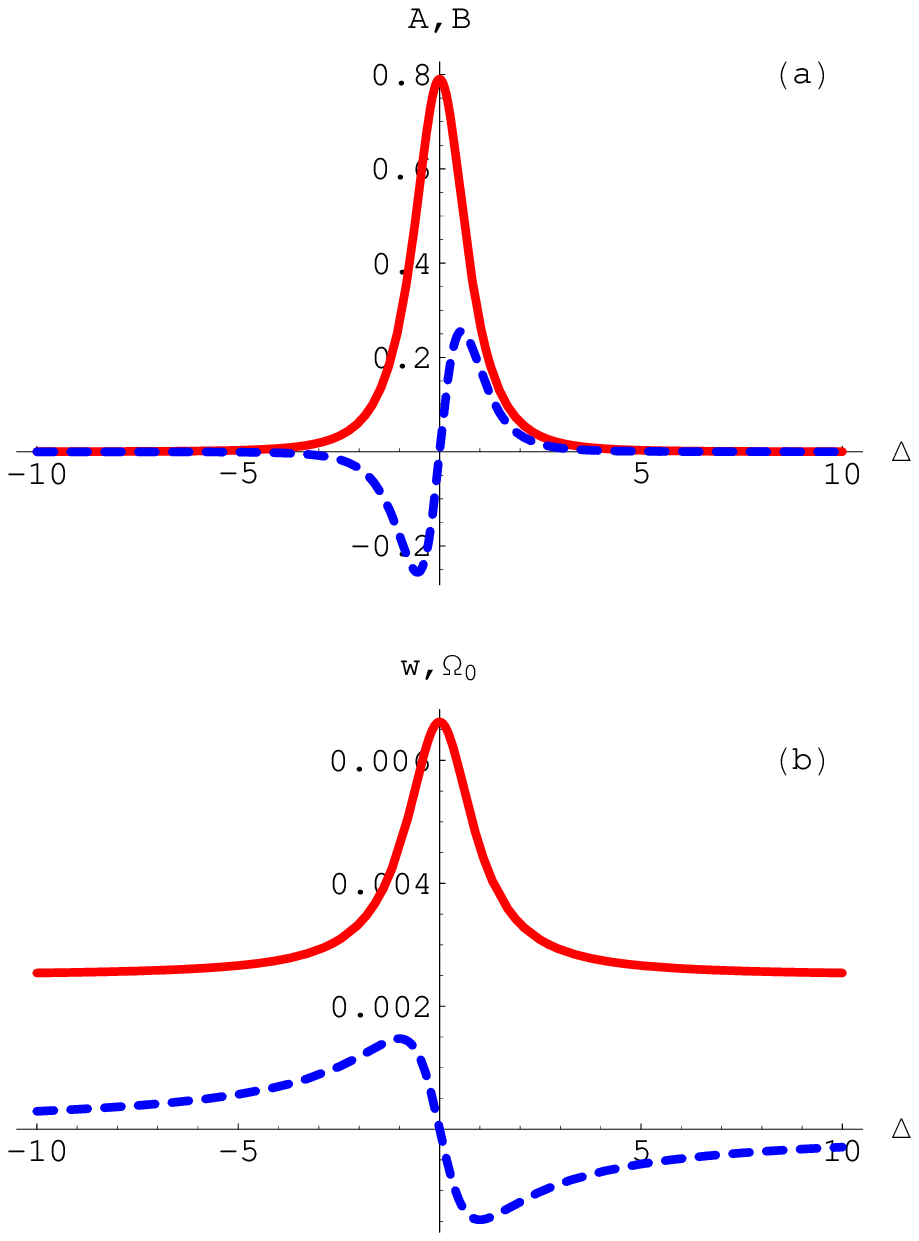}}}
\caption{The lineshape parameters versus $\Delta$. (a) $A$ (solid line) and $B$
(dashed line) in arbitrary units, and (b) $w$ (solid line) and $\Omega_0$
(dashed line) in $\gamma_{eg}$ units. The case of the pure radiative relaxation
$\gamma_1 = 0$. Other parameters $\Gamma =0.005\,\gamma_{eg}$, $\kappa =
0.1\,\gamma_{eg}$, and $\varepsilon = \pi/8$.} \label{LLparG005gk0pio8ab}
\end{figure}

It is instructive to consider the influence of the finite atom-light interaction time on the
lineshape parameters. We find that this influence becomes substantial starting from $\Gamma
\approx \,\gamma_{eg} \, S$ and it manifests mainly in the suppression of the resonance amplitudes
$A$ and $B$ especially in the wings $|\Delta| > \gamma_{eg}$. At sufficiently large $\Gamma$ the
amplitude $A$ becomes of fixed sign as shown in Fig.~\ref{LLparG005gk0pio8ab} for $\Gamma = 0.005
\,\gamma_{eg}$ in the EIA case. In a pure gas cell typically $\Gamma \approx 0.001\,-\,0.01
\gamma_{eg}$ and in order to observe, say, the Raman absorption peak at the EIT condition one needs
a buffer gas cell, where the ratio of the ground-state relaxation rate $\Gamma$ to the optical
linewidth $\gamma_{eg}$ is the more favorable $\Gamma \approx 10^{-5}\,-\,10^{-6}\,\gamma_{eg}$.
Note, the closely related results have been recently reported  in \cite{mikhailov03,budker03}.
Another strategy, more suitable for the observation of the Raman transparency dip in the EIA case,
is the use of more intense light fields.

\section{Resonance lineshape in a gas}

\begin{figure}[h]\centerline{\scalebox{1}{\includegraphics{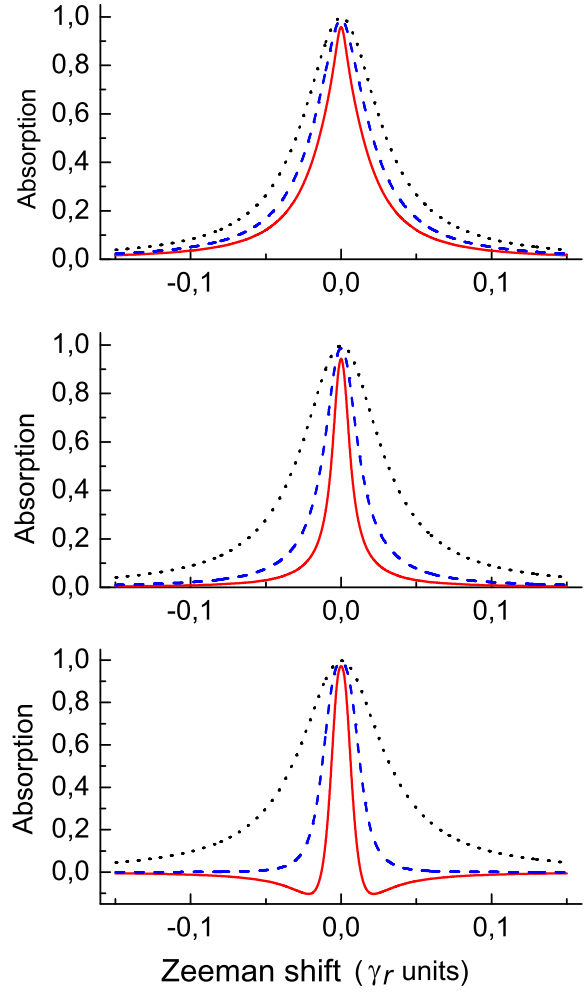}}}
\caption{The resonance lineshapes in a gas.  The case of the pure
radiative relaxation $\gamma_1 = 0$ and $\gamma_{eg} =
\gamma_r/2+\Gamma$. The total excited-state population averaged
over Maxwell velocity distribution $\langle \pi_e \rangle_v$
versus the ground-state Zeeman shift $\Omega_g$ in $\gamma_{r}$
units at different ellipticity parameters (a) $\varepsilon = 0$,
(b) $\varepsilon = \pi/10$, and (c) $\varepsilon = \pi/5$. The
Doppler width varies in each panel $k\bar{v}=0$ (dotted black
line), $k\bar{v}=\gamma_{r}$ (dashed blue line), and
$k\bar{v}=20\,\gamma_{r}$ (solid red line). Other parameters
$\delta = 0$, $\Gamma =0.001\,\gamma_{r}$, and $\kappa =
0.2\,\gamma_{r}$. For all curves the background is extracted and
all curves are normalized to the absorption in the center
$\Omega_g = 0$.} \label{lineshapesEIAkvN}
\end{figure}

In a gas with the Maxwell velocity distribution the absorption signal is
proportional to the average excited-state population:
\begin{equation}\label{ne_v}
\langle \pi_e\rangle_v=(\sqrt{\pi}\,\bar{v})^{-1}
\int_{-\infty}^{+\infty}\pi_e\,\exp\{-(v/\bar{v})^2\}\,dv\,,
\end{equation}
where the parameter $\bar{v}=(2\,k_B\,T/M)^{1/2}$, with $k_B$ the
Boltzmann constant , $T$ temperature, and $M$ mass of an atom.
This signal is a superposition of contributions of atoms with
different detunings  $\delta_v = \omega-\omega_{eg}-kv$ (due to
the Doppler shift $kv$). As is shown in the previous section, the
resonance lineshape $\pi_e(\Omega_g)$ is significantly deformed
and shifted at given nonzero detuning (see in
Fig.~\ref{lineshapesEIA}.b). As a result, the resonance lineshape
in a gas can acquire non-trivial peculiarities. For example, in
Fig.~\ref{lineshapesEIAkvN} we show the lineshapes in the EIA case
($\widehat{\gamma}_1 =0$) for different values of the ellipicity
parameter $\varepsilon$ and the average velocity $\bar{v}$. One
can see that for elliptically polarized light the resonance width
is significantly less than in the case of linear polarization.
This effect is an analog of the Doppler narrowing of the
two-photon resonance discussed in \cite{DopplerNarrowing} -
\cite{Ye}. Note, in our problem statement the circular
polarization components ${\bf e}_{\pm 1}$ with different
amplitudes at $\varepsilon \neq 0$  play the roles of ``strong''
control and ``weak'' probe fields, that is one of the conditions
for the narrowing \cite{DopplerNarrowing} - \cite{Ye}. With the
increase of $\varepsilon$ the resonance lineshape becomes
complicated -- the narrow EIA peak inside of the wider dip. This
can be viewed as a consequence of the sign reversal of the
resonance amplitude $A$ at nonzero detunings discussed above. The
sharp structures in the line center should also be noted. In
general, a detailed study of lineshape of the magneto-optical
resonances in elliptically polarized fileds in a gas is of great
interest and it can be the subject of a separate publication. Our
results in this direction will be presented elsewhere.

\section{Conclusion}
Using simple theoretical model of $1 \to 2$ closed atomic transition, the influence of the
spontaneous transfer of anisotropy on the magneto-optical absorption resonances in an elliptically
polarized field is studied. The analytical expression for the absorption for atoms with given
velocity is obtained. In the low-saturation limit, when the resonance width is much less than the
natural width, the lineshape is a generalized Lorentzian. It is shown that in the case of linearly
polarized filed ($\varepsilon=0$) or in the exact resonance ($\delta =0$) the lineshape is
symmetric, and the  resonance sign is governed by the rate of the excited-state collisional
depolarization $\gamma_1$. With an increase of $\gamma_1$ one can see a gradual transition from
the EIA type to the EIT type. In the general case of elliptical polarization $\varepsilon \neq 0$
and $\delta \neq 0$ the absorption signal is asymmetric and shifted with respect to the zero
magnetic field point.  The sign-reversal of the symmetric part of the resonance is detected at
large detuning. This effect related to the Raman scattering takes place in EIA as well as in EIT
cases. Some peculiarities of the lineshape in a gas are investigated. In particular, the Doppler
narrowing and the bimodal structure of the resonance are observed.

\begin{acknowledgments}
We thank A. Sidorov, A. Akulshin and V. Velichansky for helpful discussions. This work was
partially supported by Russian Foundation for Basic Research (grants  \#04-02-16488, \#05-02-17086,
\#05-02-16717).
\end{acknowledgments}

\newpage

\section*{List of Figure Captions}

Fig. 1. (a) Mutual orientation of the polarization ellipse and the magnetic field, and (b) the
scheme of light-induced transitions. 

\noindent Fig. 2. The lineshape parameters versus $\Delta$. (a) $A$ and (b) $B$ in arbitrary units,
and (c) $w$ and (d) $\Omega_0$ in $\gamma_{eg}$ units. The case of the pure radiative relaxation
$\gamma_1 = 0$. Other parameters $\varepsilon = \pi/8$, $\Gamma =0$, $\kappa = 0.1\,\gamma_{eg}$.

\noindent Fig. 3. The lineshape parameters versus $\Delta$. (a) $A$ and (b) $B$ in arbitrary units,
and (c) $w$ and (d) $\Omega_0$ in $\gamma_{eg}$ units. The case of the total excited-state
depolarization $\gamma_1 = 10$. Other parameters $\varepsilon = \pi/8$, $\Gamma =0$, $\kappa =
0.1\,\gamma_{eg}$.

\noindent Fig. 4. The absorption resonance lineshapes in the EIA case $\gamma_1
= 0$. (a) The total excited-state population in arbitrary units versus the
ground-state Zeeman shift in  $\gamma_{eg}$ units at $\Delta = 0$; (b) the same
but $\Delta =5\,\gamma_{eg}$. Other parameters $\varepsilon = \pi/8$, $\Gamma
=0$, $\kappa = 0.1\,\gamma_{eg}$.

\noindent Fig. 5. The lineshape parameters versus $\Delta$. (a) $A$ (solid
line) and $B$ (dashed line) in arbitrary units, and (b) $w$ (solid line) and
$\Omega_0$ (dashed line) in $\gamma_{eg}$ units. The case of the pure radiative
relaxation $\gamma_1 = 0$. Other parameters $\Gamma =0.005\,\gamma_{eg}$,
$\kappa = 0.1\,\gamma_{eg}$, and $\varepsilon = \pi/8$.

\noindent Fig. 6. The resonance lineshapes in a gas.  The case of the pure radiative relaxation
$\gamma_1 = 0$ and $\gamma_{eg} = \gamma_r/2+\Gamma$. The total excited-state population averaged
over Maxwell velocity distribution $\langle \pi_e \rangle_v$ versus the ground-state Zeeman shift
$\Omega_g$ in $\gamma_{r}$ units at different ellipticity parameters (a) $\varepsilon = 0$, (b)
$\varepsilon = \pi/10$, and (c) $\varepsilon = \pi/5$. The Doppler width varies in each panel
$k\bar{v}=0$ (dotted line), $k\bar{v}=\gamma_{r}$ (dashed line), and $k\bar{v}=20\,\gamma_{r}$
(solid line). Other parameters $\delta = 0$, $\Gamma =0.001\,\gamma_{r}$, and $\kappa =
0.2\,\gamma_{r}$. For all curves the background is subtracted and all curves are normalized to the
absorption in the center $\Omega_g = 0$.

\newpage

\end{document}